\documentclass{article}
\usepackage{spconf,graphicx,amsmath,amssymb,bm,booktabs,tabularx,multirow,color,hyperref,url}
\hypersetup{
    colorlinks=true,
    citecolor=green,
    linkcolor=red,
    urlcolor=magenta,
}
\newcolumntype{C}{>{\centering\arraybackslash}X}
\newcolumntype{L}{>{\raggedright\arraybackslash}X}
\newcolumntype{R}{>{\raggedleft\arraybackslash}X}
\makeatletter
\newcommand\footnoteref[1]{\protected@xdef\@thefnmark{\ref{#1}}\@footnotemark}
\makeatother
\title{iSTFTNet: Fast and Lightweight Mel-Spectrogram Vocoder\\
  Incorporating Inverse Short-Time Fourier Transform}

\name{Takuhiro Kaneko, Kou Tanaka, Hirokazu Kameoka, Shogo Seki}
\address{NTT Communication Science Laboratories, NTT Corporation, Japan}

\begin{document}

\maketitle

\begin{abstract}
  In recent text-to-speech synthesis and voice conversion systems, a mel-spectrogram is commonly applied as an intermediate representation, and the necessity for a mel-spectrogram vocoder is increasing. A mel-spectrogram vocoder must solve three inverse problems: recovery of the original-scale magnitude spectrogram, phase reconstruction, and frequency-to-time conversion. A typical convolutional mel-spectrogram vocoder solves these problems jointly and implicitly using a convolutional neural network, including temporal upsampling layers, when directly calculating a raw waveform. Such an approach allows skipping redundant processes during waveform synthesis (e.g., the direct reconstruction of high-dimensional original-scale spectrograms). By contrast, the approach solves all problems in a black box and cannot effectively employ the time-frequency structures existing in a mel-spectrogram. We thus propose \textit{iSTFTNet}, which replaces some output-side layers of the mel-spectrogram vocoder with the inverse short-time Fourier transform (iSTFT) after sufficiently reducing the frequency dimension using upsampling layers, reducing the computational cost from black-box modeling and avoiding redundant estimations of high-dimensional spectrograms. During our experiments, we applied our ideas to three HiFi-GAN variants and made the models faster and more lightweight with a reasonable speech quality.\footnote{\label{foot:samples}Audio samples are available at \url{https://www.kecl.ntt.co.jp/people/kaneko.takuhiro/projects/istftnet/}.}
\end{abstract}


\begin{figure}[t]
  \centerline{\includegraphics[width=\columnwidth]{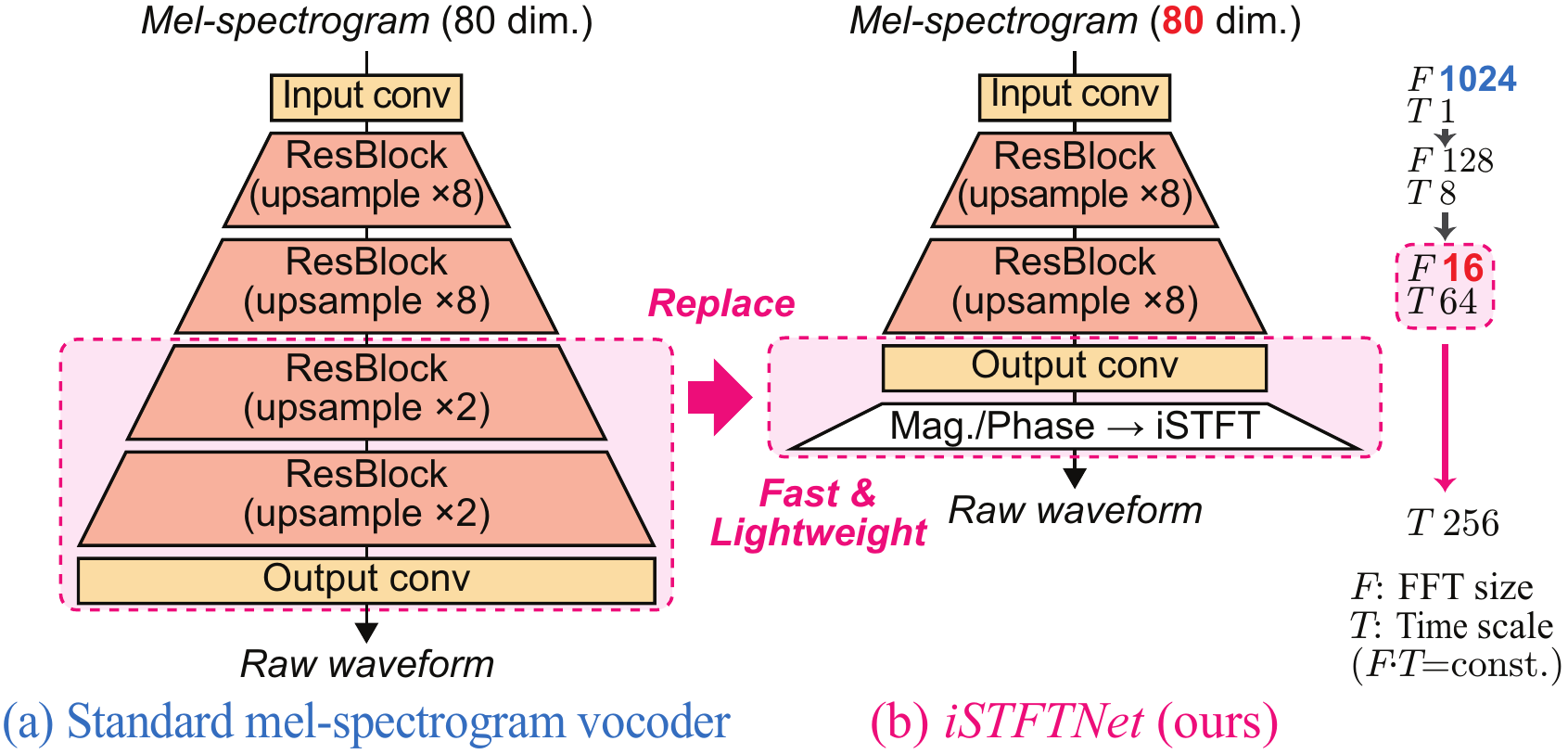}}
  \vspace{-3mm}
  \caption{Comparison of a standard convolutional mel-spectrogram vocoder and \textit{iSTFTNet} (ours).
    We propose replacing the output-side layers of the standard vocoder (a) with iSTFT (b) when the number of frequency dimensions is sufficiently small (e.g., herein, the FFT size is 16) compared to the number of dimensions of the input mel-spectrogram (80).}
  \label{fig:concept}
  \vspace{-3mm}
\end{figure}

\section{Introduction}
\label{sec:introduction}

Speech is a frequently used modality in communication, and text-to-speech (TTS) synthesis and voice conversion (VC) have been studied to eliminate human-human and human-machine boundaries.
In both TTS and VC, typical methods use a two-stage approach:
(1) The first model predicts the target intermediate representation from the text or source intermediate representations.
(2) The second step generates a raw waveform from the predicted intermediate representation.
A mel-spectrogram is widely used as an intermediate representation in recent TTS~\cite{JShenICASSP2018,WPingICLR2018,NLiAAAI2019,YRenNeurIPS2019,YRenICLR2021} and VC~\cite{KQianICML2019,TKanekoIS2020,TKanekoICASSP2021} systems owing to its compactness and expressiveness.
Consequently, the demand for a mel-spectrogram vocoder is increasing.

A mel-spectrogram vocoder must solve the following three inverse problems:
recovery of the original-scale magnitude spectrogram, phase reconstruction, and frequency-to-time conversion.
A typical convolutional mel-spectrogram vocoder (e.g., a generative adversarial network (GAN~\cite{IGoodfellowNIPS2014}) based model~\cite{KKumarNeurIPS2019,RYamamotoICASSP2020,JKongNeurIPS2020}) solves these problems jointly and implicitly using a convolutional neural network (CNN), including temporal upsampling layers, when directly calculating a raw waveform from a mel-spectrogram.
Such an approach allows omitting redundant processes during waveform synthesis, e.g., the direct reconstruction of high-dimensional original-scale spectrograms.
However, this approach solves all problems in a black box and cannot efficiently employ time-frequency structures that exist in a mel-spectrogram.

We thus propose \textit{iSTFTNet}, which replaces some output-side layers of the convolutional mel-spectrogram vocoder (Fig.~\ref{fig:concept}(a)) with well-established signal processing, particularly an inverse short-time Fourier transform (iSTFT) (Fig.~\ref{fig:concept}(b)) when the number of frequency dimensions (FFT size of 16, Fig.~\ref{fig:concept}) is sufficiently small compared to the number of dimensions of the input mel-spectrogram (80, Fig.~\ref{fig:concept}).
This reduces the computational cost from black-box modeling while avoiding redundant estimations of high-dimensional original-scale spectrograms (FFT size of 1024, Fig.~\ref{fig:concept}).
During our experiments, we applied our ideas to three HiFi-GAN variants~\cite{JKongNeurIPS2020} and made the models faster and more lightweight with a reasonable speech quality.

The rest of this paper is organized as follows.
In Section~\ref{sec:related_work}, we discuss related studies.
In Section~\ref{sec:method}, we review a typical convolutional mel-spectrogram vocoder and introduce \textit{iSTFTNet}, which is a fast and lightweight variant.
In Section~\ref{sec:experiments}, we present the experiment results.
In Section~\ref{sec:conclusion}, we provide some concluding remarks and areas of future research.

\section{Related work}
\label{sec:related_work}

Neural vocoders have been studied in speech signal processing and machine learning.
The first breakthrough was achieved using autoregressive models, including WaveNet~\cite{AOordArXiv2016} and WaveRNN~\cite{NKalchbrennerICML2018}, which achieved an impressive quality but slow inference speed owing to a sample-by-sample estimation.
Parallelizable non-autoregressive models have therefore gained attention.
For example, Parallel WaveNet~\cite{AOordICML2018} and ClariNet~\cite{WPingICLR2019} distill an autoregressive teacher model into a non-autoregressive convolutional student model.
WaveGlow~\cite{RPrengerICASSP2019} eliminates the requirement for a teacher model by incorporating Glow~\cite{DKingmaNeurIPS2018}, composed of affine coupling layers and a $1 \times 1$ invertible convolution.
WaveGrad~\cite{NChenICLR2021} and DiffWave~\cite{ZKongICLR2021}, based on diffusion probabilistic models~\cite{YSongNeurIPS2019,JHoNeurIPS2020}, apply non-autoregressive CNNs for parallel computations.
A GAN~\cite{IGoodfellowNIPS2014}-based model~\cite{KKumarNeurIPS2019,RYamamotoICASSP2020,JKongNeurIPS2020,JYangIS2020,GYangSLT2021,AMustafaICASSP2021} achieves parallelizable training and inference through noncausal convolutions.
As described, CNNs with temporal upsampling layers, shown in Fig.~\ref{fig:concept}(a), have been commonly used in recent mel-spectrogram vocoders.
Thus, beyond the HiFi-GANs~\cite{JKongNeurIPS2020} used in our experiments, our ideas are general and can be applied to other models.

The use of iSTFT for neural speech synthesis was previously introduced~\cite{KOyamadaEUSIPCO2018,PNeekharaIS2019,AGritsenkoNeurIPS2020} (including our own early attempts~\cite{KOyamadaEUSIPCO2018}).
As the main difference between the previous models and \textit{iSTFTNet}, the former requires a high-capacity or high-computational model (e.g., 12 residual blocks with 2048 channels~\cite{AGritsenkoNeurIPS2020} and 2D CNNs~\cite{KOyamadaEUSIPCO2018,PNeekharaIS2019}) because they aim to reconstruct the original-scale spectrograms without changing the time scale.
By contrast, \textit{iSTFTNet} employs a hybrid approach in which iSTFT is applied after some upsampling processes (Fig.~\ref{fig:concept}(b)).
This allows a reasonable performance using a low-capacity model (e.g., 1D CNNs, commonly used in typical GAN vocoders~\cite{KKumarNeurIPS2019,RYamamotoICASSP2020,JKongNeurIPS2020,JYangIS2020,GYangSLT2021,AMustafaICASSP2021}).

\section{Method}
\label{sec:method}

\subsection{Convolutional mel-spectrogram vocoder}
\label{subsec:melspectrogram_vocoder}

\begin{figure}[t]
  \centerline{\includegraphics[width=\columnwidth]{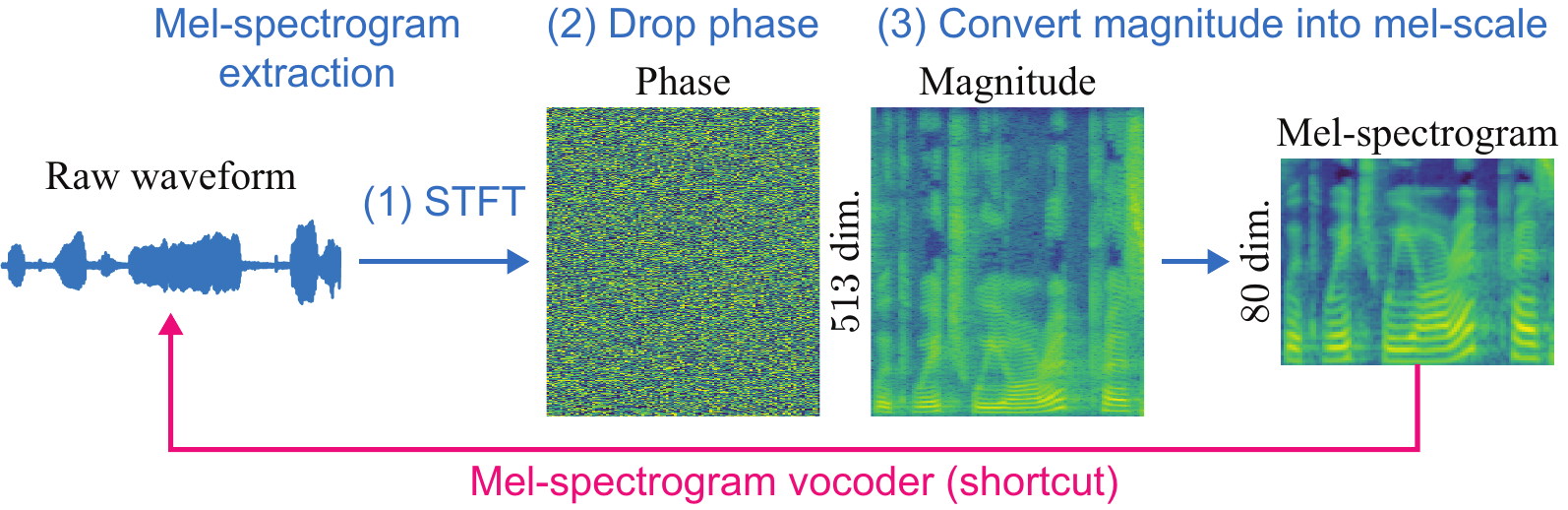}}
  \vspace{-3mm}
  \caption{Processing flows of mel-spectrogram extraction (light blue) and mel-spectrogram vocoder (pink)}
  \label{fig:processing_flows}
  \vspace{-3mm}
\end{figure}

As shown in Fig.~\ref{fig:processing_flows}, the mel-spectrogram is extracted from the raw waveform as follows:
(1) The magnitude and phase spectrograms are extracted from the raw waveform using a short-time Fourier transform (STFT).
(2) The phase spectrogram is dropped.
(3) The magnitude spectrogram is converted into a mel-scale.
Because a mel-spectrogram vocoder is aimed at an inverse process, three inverse problems must be solved:
(3') recovery of the original-scale magnitude spectrogram;
(2') phase reconstruction; and
(1') frequency-to-time conversion.

A typical convolutional mel-spectrogram vocoder solves these problems jointly and implicitly using a CNN, including temporal upsampling layers, while directly calculating a raw waveform from a mel-spectrogram.
This approach allows redundant processes (e.g., the direct reconstruction of high-dimensional original-scale magnitude and phase spectrograms) to be skipped during waveform synthesis.
This allows a convolutional mel-spectrogram vocoder to solve the aforementioned problems with a low-capacity model.
For example, HiFi-GAN V2~\cite{JKongNeurIPS2020} achieves a good performance using only 1D convolutions of channels smaller than 128, despite being smaller than the original-scale spectrogram dimensions (i.e., 513, Fig.~\ref{fig:processing_flows}).

\begin{figure*}[t]
  \centerline{\includegraphics[width=\textwidth]{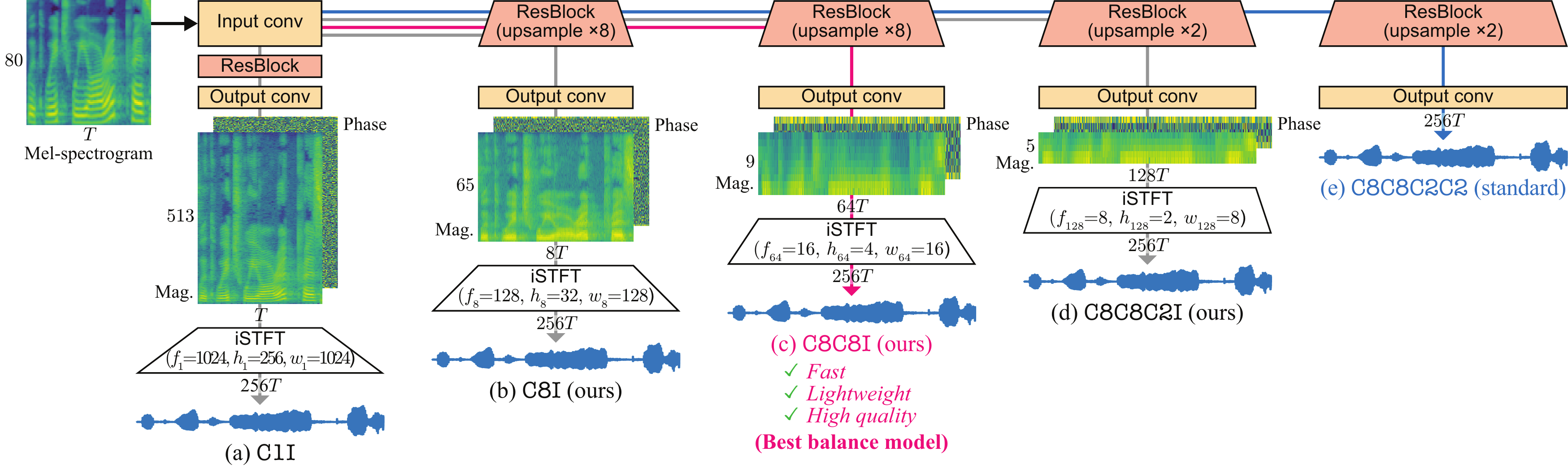}}
  \vspace{-3mm}
  \caption{Architectures of \textit{iSTFTNets} (b)--(d) and a standard convolutional mel-spectrogram vocoder (e).
    The model is denoted as {\tt C$x$$\dots$(I)}, where {\tt C$x$} indicates the use of a residual block (ResBlock)~\cite{KHeCVPR2016} with an $\times x$ upsampling layer and {\tt I} indicates the use of iSTFT.
    Here, the input 80-dimensional mel-spectrogram was extracted from a 22.05-kHz waveform using STFT with an FFT size of 1024, hop length of 256, and window length of 1024.}
  \label{fig:architecture}
  \vspace{-3mm}
\end{figure*}

\subsection{iSTFTNet: Fast and lightweight vocoder with iSTFT}
\label{subsec:istftnet}

Black-box modeling is useful for discovering potential shortcuts.
However, we cannot effectively employ the time-frequency structures existing in the mel-spectrogram despite providing a hint for solving inverse problems.

Thus, we propose \textit{iSTFTNet}, which employs time and frequency structures explicitly using iSTFT after sufficiently reducing the frequency dimension using some upsampling layers, as shown in Fig.~\ref{fig:architecture}(b)--(d).
Here, we utilize the characteristics of STFT, that is, the trade-off between the time and frequency resolution.
More precisely, when the iSTFT required after $s \times$ upsampling is represented as $\text{iSTFT}(f_s, h_s, w_s)$, where $f_s$, $h_s$, and $w_s$ indicate the FFT size, hop length, and window length, respectively; $\text{iSTFT}(f_s, h_s, w_s)$ can be calculated using the parameters of iSTFT required for the original-scale spectrogram, $\text{iSTFT}(f_1, h_1, w_1)$:
\begin{flalign}
  \label{eq:istft}
  \text{iSTFT} (f_s, h_s, w_s)
  = \text{iSTFT} \left( \frac{f_1}{s}, \frac{h_1}{s}, \frac{w_1}{s} \right),
\end{flalign}
where we utilize the aforementioned STFT characteristic, that is, $f_1 \cdot 1 = f_s \cdot s = \text{constant}$.
This equation means that we can reduce the frequency dimensions by increasing $s$.
As shown in Fig.~\ref{fig:architecture}, we can simplify the structure in the frequency direction by increasing the number of upsamples.
In Section~\ref{subsec:evaluation}, we empirically found that simplification through more than two upsamples (Fig.~\ref{fig:architecture}(c) or (d)) is essential for a faster and more lightweight model to achieve a reasonable quality.

\subsection{Implementation}
\label{subsec:implementation}

\textit{iSTFTNets} (Fig.~\ref{fig:architecture}(b)--(d)) have the mostly same network architecture as the baseline (Fig.~\ref{fig:architecture}(e)).
Hence, when a reliable convolutional mel-spectrogram vocoder is obtained, it is easy to incorporate the concept of \textit{iSTFTNet}.
However, three minor but essential modifications are required:
(i) The output channels of the final convolutional layer should be changed from 1 to $(f_s / 2 + 1) \times 2$ to generate magnitude and phase spectrograms instead of a raw waveform.
(ii) Exponential and sine activation functions should be applied to the output of (i) when calculating the magnitude and phase spectrograms, respectively.
(iii) A raw waveform should be generated from the magnitude and phase spectrograms using iSTFT (Eq.~(\ref{eq:istft})).
For (ii), we use an exponential activation function because the required magnitude spectrogram uses a linear scale, whereas the input mel-spectrogram uses a log scale, and we apply a sine activation function to represent the periodic characteristics of the phase spectrogram.

\section{Experiments}
\label{sec:experiments}

\subsection{Experiment setup}
\label{subsec:experiment_setup}

\noindent\textbf{Dataset.}
We used the LJSpeech dataset~\cite{ljspeech17}, consisting of 13,100 audio clips (24 h) of a female speaker.
Here, 12,600, 250, and 250 utterances were used for the training, validation, and evaluation, respectively.
The audio clips were sampled at 22.05 kHz, and 80-dimensional log-mel spectrograms were extracted with an FFT size of 1024, hop length of 256, and window length of 1024.

\smallskip\noindent\textbf{Network architectures.}
We applied our ideas to three HiFi-GAN variants~\cite{JKongNeurIPS2020} (high-quality (V1), light (V2), and carefully tuned (V3) variants).
We implemented them based on an open-source code\footnote{\label{foot:parallel_wavegan}\url{https://github.com/kan-bayashi/ParallelWaveGAN}} for fair comparison with the various synthesis speeches provided.
As mentioned in Section~\ref{subsec:implementation}, with the exception of the three modifications described above, we used the same architectures as the baselines.

\smallskip\noindent\textbf{Training settings.}
We trained the models using the HiFi-GAN configuration provided in the open-source code,\footnoteref{foot:parallel_wavegan} the parameters of which were tuned for stable training across various datasets.
We trained the model for 2.5M iterations using the Adam optimizer~\cite{DPKingmaICLR2015} with an initial learning rate of 0.0002, and momentum terms $\beta_1$ and $\beta_2$ of 0.5 and 0.9, respectively.
For the loss function, we used a combination of least squares GAN~\cite{XMaoICCV2017}, mel-spectrogram~\cite{JKongNeurIPS2020}, and feature matching~\cite{ALarsenICML2016,KKumarNeurIPS2019} losses.

\subsection{Evaluation}
\label{subsec:evaluation}

We conducted a mean opinion score (MOS) test to evaluate the perceptual quality, randomly selecting 20 utterances from the evaluation set and using the ground truth mel-spectrograms of the utterances as the vocoder input.
This test was conducted online with 16 listeners.
Audio samples are available from the link\footnoteref{foot:samples} presented on the first page.
As an objective metric, we used the conditional Fr\'{e}chet wav2vec distance (\textit{cFW2VD}), which measures the distance between real and generative distributions in a wav2vec 2.0~\cite{ABaevskiNeurIPS2020} feature space conditioned on text information.
This is conceptually similar to the Fr\'{e}chet inception distance (FID)~\cite{MHeuselNIPS2017} and Fr\'{e}chet DeepSpeech distance (FDSD)~\cite{MBinkowskiICLR2020}, which measure the perceptual quality of images and speeches, respectively.
We used cFW2VD instead of conditional FDSD (cFDSD)~\cite{MBinkowskiICLR2020} to evaluate the raw waveform directly without converting it into a power spectrogram, as required in cFDSD.
We found that MOS has a higher correlation with cFW2VD than with cFDSD (Spearman's rank correlation of -0.93 and -0.83, respectively).\footnote{We provide the detailed analysis in Appendix~\ref{subsec:mos_cfw2vd_cfdsd}.}
In cFW2VD, the smaller the value is, the better the perceptual quality.
Table~\ref{tab:result_vocoder} shows the results, along with the inference speed and model size.
We examined the approach from three perspectives.

\begin{table}[t]
  \caption{Comparison of MOS with 95\% confidence intervals, cFW2VD, inference speed, and model size.
    The inference speed (relative speed compared to real time) using a GPU was calculated on a single NVIDIA V100 GPU, and the speed using a CPU was computed on a MacBook Pro laptop (2.7-GHz Intel Core i7).
    We report the average score over the utterances in the evaluation set.
    The model identifier (e.g., {\tt C8C8I}) is shown in Fig.~\ref{fig:architecture}.
    The numbers in () indicate the rates (\%) compared with the baselines (V1, V2, or V3).
    The underlined models are \textit{iSTFTNets} (fast and lightweight models).}
  \label{tab:result_vocoder}
  \vspace{1mm}
  \newcommand{\spm}[1]{{\tiny$\pm$#1}}
  \newcommand{\rate}[1]{{\tiny({#1})}}
  \setlength{\tabcolsep}{1pt}
  \renewcommand{\arraystretch}{0.94}
  \centering
  \scriptsize{
  \begin{tabularx}{\columnwidth}{clccrrr}
    \toprule
    \multirow{2}{*}{\textbf{No.}} & \multicolumn{1}{c}{\multirow{2}{*}{\textbf{Model}}} & \multirow{2}{*}{\textbf{MOS}$\uparrow$} & \multirow{2}{*}{\textbf{cFW2VD}$\downarrow$} & \multicolumn{1}{c}{\textbf{Speed}$\uparrow$} & \multicolumn{1}{c}{\textbf{Speed}$\uparrow$} & \multicolumn{1}{c}{\# \textbf{Param}$\downarrow$}
    \\
    & & & & \multicolumn{1}{c}{\textbf{(GPU)}} & \multicolumn{1}{c}{\textbf{(CPU)}} & \multicolumn{1}{c}{\textbf{(M)}}
    \\ \midrule
    \!1\! & Ground truth
    & 4.46 \spm{0.14} & \multicolumn{1}{c}{--} & \multicolumn{1}{c}{--} & \multicolumn{1}{c}{--} & \multicolumn{1}{c}{--}
    \\ \midrule
    \!2\! & V1 (original)\tiny{~\cite{JKongNeurIPS2020}}
    & 4.22 \spm{0.17} & 0.020 & $\times$143.59 \rate{100}\: & $\times$1.34 \rate{100}\: & 13.94 \rate{100}\:\:\:\:
    \\
    \!3\! & \underline{V1-{\tt C8C8C2I}}
    & 4.22 \spm{0.17} & 0.018 & $\times$179.42 \rate{125}\: & $\times$1.63 \rate{122}\: & 13.80 \rate{\; 99}\:\:\:\:
    \\
    \!4\! & \underline{V1-{\tt C8C8I}}
    & 4.26 \spm{0.17} & 0.020 & $\times$245.68 \rate{171}\: & $\times$2.33 \rate{174}\: & 13.26 \rate{\; 95}\:\:\:\:
    \\
    \!5\! & \underline{V1-{\tt C8I}}
    & 3.32 \spm{0.22} & 0.073 & $\times$609.43 \rate{424}\: & $\times$7.57 \rate{565}\: & 10.89 \rate{\; 78}\:\:\:\:
    \\
    \!6\! & V1-{\tt C8C1I}
    & 3.82 \spm{0.17} & 0.033 & $\times$326.39 \rate{227}\: & $\times$3.97 \rate{296}\: & 19.15 \rate{137}\:\:\:\:
    \\ \midrule
    \!7\! & V2 (original)\tiny{~\cite{JKongNeurIPS2020}}
    & 3.91 \spm{0.17} & 0.046 & $\times$624.47 \rate{100}\: & $\times$10.39 \rate{100}\: & 0.93 \rate{100}\:\:\:\:
    \\
    \!8\! & \underline{V2-{\tt C8C8C2I}}
    & 3.98 \spm{0.17} & 0.038 & $\times$732.96 \rate{117}\: & $\times$13.34 \rate{128}\: & 0.92 \rate{\; 99}\:\:\:\:
    \\
    \!9\! & \underline{V2-{\tt C8C8I}}
    & 3.95 \spm{0.16} & 0.042 & $\times$1025.46 \rate{164}\: & $\times$20.37 \rate{196}\: & 0.89 \rate{\; 96}\:\:\:\:
    \\
    \!10\! & \underline{V2-{\tt C8I}}
    & 3.21 \spm{0.20} & 0.096 & $\times$1720.91 \rate{276}\: & $\times$68.05 \rate{655}\: & 0.78 \rate{\; 84}\:\:\:\:
    \\
    \!11\! & V2-{\tt C8C1I}
    & 3.44 \spm{0.20} & 0.071 & $\times$1081.37 \rate{173}\: & $\times$39.14 \rate{377}\: & 1.30 \rate{140}\:\:\:\:
    \\ \midrule
    \!12\! & V3 (original)\tiny{~\cite{JKongNeurIPS2020}}
    & 3.78 \spm{0.16} & 0.052 & $\times$933.06 \rate{100}\: & $\times$10.40 \rate{100}\: & 1.46 \rate{100}\:\:\:\:
    \\
    \!13\! & \underline{V3-{\tt C8C8I}}
    & 3.41 \spm{0.19} & 0.055 & $\times$1517.70 \rate{163}\: & $\times$21.48 \rate{206}\: & 1.42 \rate{\; 97}\:\:\:\:
    \\
    \!14\! & \underline{V3-{\tt C8I}}
    & 2.89 \spm{0.17} & 0.156 & $\times$2481.87 \rate{266}\: & $\times$66.83 \rate{642}\: & 1.28 \rate{\; 87}\:\:\:\:
    \\
    \!15\! & V3-{\tt C8C1I}
    & 2.82 \spm{0.21} & 0.116 & $\times$1925.15 \rate{206}\: & $\times$41.16 \rate{396}\: & 1.77 \rate{121}\:\:\:\:
    \\ \midrule
    \!16\! & MB-MelGAN\tiny{~\cite{GYangSLT2021}}
    & 3.54 \spm{0.21} & 0.078 & $\times$1070.95 \:\:\:\:\:\:\:\: & $\times$17.95 \:\:\:\:\:\:\:\: & 2.54 \:\:\:\:\:\:\:\:\:\:\,
    \\
    \!17\! & PWG\tiny{~\cite{RYamamotoICASSP2020}}
    & 3.47 \spm{0.21} & 0.066 & $\times$79.71 \:\:\:\:\:\:\:\: & $\times$0.70 \:\:\:\:\:\:\:\: & 1.35 \:\:\:\:\:\:\:\:\:\:\,
    \\ \bottomrule    
  \end{tabularx}
  }
  \vspace{-3mm}
\end{table}

\smallskip\noindent\textbf{(1) How many layers should be replaced with iSTFT?}
The vocoder is improved by replacing its output-side layers with a faster and more lightweight iSTFT.
To determine the number of layers to be replaced, we investigated the performance differences between the models shown in Fig.~\ref{fig:architecture}.
The corresponding results are listed in Table~\ref{tab:result_vocoder} (Nos. 2--5, 7--10, and 12--14).\footnote{{\tt C8C8C2I} was not used for V3 because the network of V3 is {\tt C8C8C4} and does not have a fourth upsampling layer, differently from V1 and V2.}
As expected, the inference speeds up, and the model size decreases with more replaced layers.
For the MOS, we found that {\tt C8I} performs worse than the original in all cases; however, {\tt C8C8I} and {\tt C8C8C2I} for V1 and V2 were comparable to the original.
For V3 only, the performance decreases when {\tt C8C8I} is used.
This is because V3 is carefully tuned to reduce the number of layers and loses its generality.
However, note that despite the performance decrease, V3-{\tt C8C8I} is still comparable with Parallel WaveGAN (PWG)~\cite{RYamamotoICASSP2020} (Table~\ref{tab:result_vocoder} (No.~17)), while improving the inference speed.

\smallskip\noindent\textbf{(2) Necessity of combining upsampling and iSTFT.}
The numbers of both the upsampling layers and residual blocks differ between {\tt C8I} and {\tt C8C8I}, as shown in Fig.~\ref{fig:architecture}(b) and (c).
To solve this problem, we examined the performance of {\tt C8C1I}\footnote{The model size of {\tt C8C1I} is larger than that of {\tt C8C8I} because in the latter, the channels are halved in the second ResBlock after upsampling, whereas in the former, this is not conducted owing to the absence of upsampling.} which applies one upsampling but uses two residual blocks, similar to {\tt C8C8I}.
The corresponding results are presented in Table~\ref{tab:result_vocoder} (Nos. 6, 11, and 15).
We found that {\tt C8C1I} is still worse than {\tt C8C8I}, indicating that reducing the frequency dimension using upsampling is essential for obtaining a reasonable quality when applying iSTFT without significant changes to the network architecture.\footnote{We also examined non-upsampling models (particularly, {\tt C1I} (Fig.~\ref{fig:architecture}(a)) and {\tt C1C1I}). Finding that they suffer from training difficulties with a significantly lower speech quality, we omitted them from the experiments.}

\smallskip\noindent\textbf{(3) Comparison with fastest baseline.}
One of the fastest GAN vocoders is multi-band (MB) MelGAN~\cite{GYangSLT2021}, which increases the speed of MelGAN~\cite{KKumarNeurIPS2019} by changing the synthesis target from a full-band signal to lower-resolution sub-band signals~\cite{CYuIS2020}.
To examine the validity of this speed, we compared our models with MB-MelGAN.
The corresponding results are presented in Table~\ref{tab:result_vocoder} (No. 16).
Here, V2-{\tt C8C8I} outperformed MB-MelGAN for MOS, reducing the model size and achieving a comparable speed.
Note that the multi-band formulation and iSTFT are orthogonal and compatible, and iSTFT can be incorporated into MB-MelGAN using $\text{iSTFT} \left( \frac{f_1}{sb}, \frac{h_1}{sb}, \frac{w_1}{sb} \right)$, where $b$ is the number of sub-bands.
This approach remains for future research.\footnote{As another difference between \textit{iSTFTNet} and MB-MelGAN, MB-MelGAN requires an additional sub-band STFT loss for stable training, whereas \textit{iSTFTNet} does not.}

\subsection{Application to text-to-speech synthesis}
\label{subsec:application_to_text_to_speech_synthesis}

Next, we examine the effectiveness of our approach when applied to TTS synthesis, focusing on the difference in performance between V1 and V1-{\tt C8C8I} when combined with Conformer-FS2~\cite{PGuoICASSP2021} (a combination of Conformer~\cite{AGulatiIS2020} and FastSpeech 2 (FS2)~\cite{YRenICLR2021}).
Following~\cite{JKongNeurIPS2020}, which shows the utility of fine-tuning on HiFi-GAN, we fine-tuned the combined models for 300k iterations in an end-to-end manner after training each model.
We applied an open-source code~\cite{THayashiICASSP2020}\footnote{\label{foot:espnet}\url{https://github.com/espnet/espnet}} for fair comparison with other speech samples provided.
We conducted a MOS test to evaluate the perceptual quality by randomly selecting 20 utterances from the evaluation set.
This test was conducted online with 12 listeners.
Audio samples are available from the link\footnoteref{foot:samples} presented on the first page.

Table~\ref{tab:result_tts} summarizes the results.
We found that V1-{\tt C8C8I} not only achieves a comparable or better performance than V1 and Conformer-FS2, but also is comparable with the ground truth.
These results indicate that \textit{iSTFTNet} does not compromise the speech quality, even for TTS synthesis.

\begin{table}[t]
  \caption{Comparison of MOS with 95\% confidence intervals and cFW2VD for TTS synthesis}
  \label{tab:result_tts}
  \vspace{1mm}
  \newcommand{\spm}[1]{{\tiny$\pm$#1}}
  \setlength{\tabcolsep}{8pt}
  \renewcommand{\arraystretch}{0.94}
  \centering
  \scriptsize{
  \begin{tabularx}{0.9\columnwidth}{clcc}
    \toprule
    \textbf{No.} & \multicolumn{1}{c}{\textbf{Model}} & \textbf{MOS}$\uparrow$ & \textbf{cFW2VD}$\downarrow$
    \\ \midrule
    1 & Ground truth
    & 4.32 \spm{0.10} & --
    \\ \midrule
    2 & Conformer-FS2 + V1
    & 4.09 \spm{0.12} & 0.216
    \\
    3 & Conformer-FS2 + \underline{V1-{\tt C8C8I}}
    & 4.25 \spm{0.11} & 0.214
    \\ \midrule
    4 & Conformer-FS2~\cite{PGuoICASSP2021}
    & 3.66 \spm{0.15} & 0.242
    \\ \bottomrule    
  \end{tabularx}
  }
  \vspace{-3mm}
\end{table}

\section{Conclusion}
\label{sec:conclusion}

To employ time-frequency structures in a mel-spectrogram while avoiding redundant estimations of high-dimensional original-scale spectrograms, we propose \textit{iSTFTNet}, replacing the output-side layers of a convolutional mel-spectrogram vocoder with iSTFT after reducing the frequency dimension using upsampling layers.
The experiment results demonstrate that we can make the models faster and more lightweight using iSTFT, and that upsampling processes are essential for obtaining a reasonable quality.
As discussed in Section~\ref{sec:related_work}, CNNs with upsampling layers are used by various vocoders beyond GAN-based versions.
Hence, applying our idea to such vocoders will be interesting.
We also concurrently investigate the utility of the inverse fast Fourier transform for a recurrent neural vocoder, and plan to examine the difference in performance in future studies to further validate the utility of the inverse Fourier transform.

\smallskip\noindent\textbf{Acknowledgements.}
This work was partially supported by JST CREST Grant Number JPMJCR19A3, Japan.

\clearpage
\bibliographystyle{IEEEbib}
{\footnotesize\bibliography{refs}}

\begin{thebibliography}{10}

\bibitem{JShenICASSP2018}
Jonathan Shen, Ruoming Pang, Ron~J. Weiss, Mike Schuster, Navdeep Jaitly,
  Zongheng Yang, Zhifeng Chen, Yu~Zhang, Yuxuan Wang, Rj~Skerrv-Ryan, Rif~A.
  Saurous, Yannis Agiomyrgiannakis, and Yonghui Wu,
\newblock ``Natural {TTS} synthesis by conditioning {WaveNet} on mel
  spectrogram predictions,''
\newblock in {\em Proc. ICASSP}, 2018, pp. 4779--4783.

\bibitem{WPingICLR2018}
Wei Ping, Kainan Peng, Andrew Gibiansky, Sercan~O. Arik, Ajay Kannan, Sharan
  Narang, Jonathan Raiman, and John Miller,
\newblock ``{Deep Voice 3}: Scaling text-to-speech with convolutional sequence
  learning,''
\newblock in {\em Proc. ICLR}, 2018.

\bibitem{NLiAAAI2019}
Naihan Li, Shujie Liu, Yanqing Liu, Sheng Zhao, and Ming Liu,
\newblock ``Neural speech synthesis with transformer network,''
\newblock in {\em Proc. AAAI}, 2019, pp. 6706--6713.

\bibitem{YRenNeurIPS2019}
Yi~Ren, Yangjun Ruan, Xu~Tan, Tao Qin, Sheng Zhao, Zhou Zhao, and Tie-Yan Liu,
\newblock ``{FastSpeech}: Fast, robust and controllable text to speech,''
\newblock in {\em Proc. NeurIPS}, 2019.

\bibitem{YRenICLR2021}
Yi~Ren, Chenxu Hu, Xu~Tan, Tao Qin, Sheng Zhao, Zhou Zhao, and Tie-Yan Liu,
\newblock ``{FastSpeech 2}: Fast and high-quality end-to-end text to speech,''
\newblock in {\em Proc. ICLR}, 2021.

\bibitem{KQianICML2019}
Kaizhi Qian, Yang Zhang, Shiyu Chang, Xuesong Yang, and Mark Hasegawa-Johnson,
\newblock ``{Auto-VC}: Zero-shot voice style transfer with only autoencoder
  loss,''
\newblock in {\em Proc. ICML}, 2019, pp. 5210--5219.

\bibitem{TKanekoIS2020}
Takuhiro Kaneko, Hirokazu Kameoka, Kou Tanaka, and Nobukatsu Hojo,
\newblock ``{CycleGAN-VC3}: Examining and improving {CycleGAN-VCs} for
  mel-spectrogram conversion,''
\newblock in {\em Proc. Interspeech}, 2020, pp. 2017--2021.

\bibitem{TKanekoICASSP2021}
Takuhiro Kaneko, Hirokazu Kameoka, Kou Tanaka, and Nobukatsu Hojo,
\newblock ``{MaskCycleGAN-VC}: Learning non-parallel voice conversion with
  filling in frames,''
\newblock in {\em Proc. ICASSP}, 2021, pp. 5919--5923.

\bibitem{IGoodfellowNIPS2014}
Ian Goodfellow, Jean Pouget-Abadie, Mehdi Mirza, Bing Xu, David Warde-Farley,
  Sherjil Ozair, Aaron Courville, and Yoshua Bengio,
\newblock ``Generative adversarial nets,''
\newblock in {\em Proc. NIPS}, 2014, pp. 2672--2680.

\bibitem{KKumarNeurIPS2019}
Kundan Kumar, Rithesh Kumar, Thibault de~Boissiere, Lucas Gestin, Wei~Zhen
  Teoh, Jose Sotelo, Alexandre de~Br{\'e}bisson, Yoshua Bengio, and Aaron
  Courville,
\newblock ``{MelGAN}: Generative adversarial networks for conditional waveform
  synthesis,''
\newblock in {\em Proc. NeurIPS}, 2019.

\bibitem{RYamamotoICASSP2020}
Ryuichi Yamamoto, Eunwoo Song, and Jae-Min Kim,
\newblock ``{Parallel WaveGAN}: A fast waveform generation model based on
  generative adversarial networks with multi-resolution spectrogram,''
\newblock in {\em Proc. ICASSP}, 2020, pp. 6199--6203.

\bibitem{JKongNeurIPS2020}
Jungil Kong, Jaehyeon Kim, and Jaekyoung Bae,
\newblock ``{HiFi-GAN}: Generative adversarial networks for efficient and high
  fidelity speech synthesis,''
\newblock in {\em Proc. NeurIPS}, 2020.

\bibitem{AOordArXiv2016}
A\"aron van~den Oord, Sander Dieleman, Heiga Zen, Karen Simonyan, Oriol
  Vinyals, Alex Graves, Nal Kalchbrenner, Andrew Senior, and Koray Kavukcuoglu,
\newblock ``{WaveNet}: A generative model for raw audio,''
\newblock {\em arXiv preprint arXiv:1609.03499}, 2016.

\bibitem{NKalchbrennerICML2018}
Nal Kalchbrenner, Erich Elsen, Karen Simonyan, Seb Noury, Norman Casagrande,
  Edward Lockhart, Florian Stimberg, A\"aron van~den Oord, Sander Dieleman, and
  Koray Kavukcuoglu,
\newblock ``Efficient neural audio synthesis,''
\newblock in {\em Proc. ICML}, 2018, pp. 2410--2419.

\bibitem{AOordICML2018}
A\"aron van~den Oord, Yazhe Li, Igor Babuschkin, Karen Simonyan, Oriol Vinyals,
  Koray Kavukcuoglu, George van~den Driessche, Edward Lockhart, Luis Cobo,
  Florian Stimberg, Norman Casagrande, Dominik Grewe, Seb Noury, Sander
  Dieleman, Erich Elsen, Nal Kalchbrenner, Heiga Zen, Alex Graves, Helen King,
  Tom Walters, Dan Belov, and Demis Hassabis,
\newblock ``Parallel {W}ave{N}et: Fast high-fidelity speech synthesis,''
\newblock in {\em Proc. ICML}, 2018, pp. 3918--3926.

\bibitem{WPingICLR2019}
Wei Ping, Kainan Peng, and Jitong Chen,
\newblock ``{ClariNet}: Parallel wave generation in end-to-end
  text-to-speech,''
\newblock in {\em Proc. ICLR}, 2019.

\bibitem{RPrengerICASSP2019}
Ryan Prenger, Rafael Valle, and Bryan Catanzaro,
\newblock ``{WaveGlow}: A flow-based generative network for speech synthesis,''
\newblock in {\em Proc. ICASSP}, 2019, pp. 3617--3621.

\bibitem{DKingmaNeurIPS2018}
Diederik~P. Kingma and Prafulla Dhariwal,
\newblock ``{Glow}: Generative flow with invertible $1\times1$ convolutions,''
\newblock in {\em Proc. NeurIPS}, 2018.

\bibitem{NChenICLR2021}
Nanxin Chen, Yu~Zhang, Heiga Zen, Ron~J. Weiss, Mohammad Norouzi, and William
  Chan,
\newblock ``{WaveGrad}: Estimating gradients for waveform generation,''
\newblock in {\em Proc. ICLR}, 2020.

\bibitem{ZKongICLR2021}
Zhifeng Kong, Wei Ping, Jiaji Huang, Kexin Zhao, and Bryan Catanzaro,
\newblock ``{DiffWave}: A versatile diffusion model for audio synthesis,''
\newblock in {\em Proc. ICLR}, 2021.

\bibitem{YSongNeurIPS2019}
Yang Song and Stefano Ermon,
\newblock ``Generative modeling by estimating gradients of the data
  distribution,''
\newblock in {\em Proc. NeurIPS}, 2019.

\bibitem{JHoNeurIPS2020}
Jonathan Ho, Ajay Jain, and Pieter Abbeel,
\newblock ``Denoising diffusion probabilistic models,''
\newblock in {\em Proc. NeurIPS}, 2020.

\bibitem{JYangIS2020}
Jinhyeok Yang, Junmo Lee, Youngik Kim, Hoonyoung Cho, and Injung Kim,
\newblock ``{VocGAN}: A high-fidelity real-time vocoder with a
  hierarchically-nested adversarial network,''
\newblock in {\em Proc. Interspeech}, 2020, pp. 200--204.

\bibitem{GYangSLT2021}
Geng Yang, Shan Yang, Kai Liu, Peng Fang, Wei Chen, and Lei Xie,
\newblock ``{Multi-band MelGAN}: Faster waveform generation for high-quality
  text-to-speech,''
\newblock in {\em Proc. SLT}, 2021, pp. 492--498.

\bibitem{AMustafaICASSP2021}
Ahmed Mustafa, Nicola Pia, and Guillaume Fuchs,
\newblock ``{StyleMelGAN}: An efficient high-fidelity adversarial vocoder with
  temporal adaptive normalization,''
\newblock in {\em Proc. ICASSP}, 2021, pp. 6034--6038.

\bibitem{KOyamadaEUSIPCO2018}
Keisuke Oyamada, Hirokazu Kameoka, Takuhiro Kaneko, Kou Tanaka, Nobukatsu Hojo,
  and Hiroyasu Ando,
\newblock ``Generative adversarial network-based approach to signal
  reconstruction from magnitude spectrogram,''
\newblock in {\em Proc. EUSIPCO}, 2018, pp. 2514--2518.

\bibitem{PNeekharaIS2019}
Paarth Neekhara, Chris Donahue, Miller Puckette, Shlomo Dubnov, and Julian
  McAuley,
\newblock ``Expediting {TTS} synthesis with adversarial vocoding,''
\newblock in {\em Proc. Interspeech}, 2019, pp. 186--190.

\bibitem{AGritsenkoNeurIPS2020}
Alexey~A. Gritsenko, Tim Salimans, Rianne van~den Berg, Jasper Snoek, and Nal
  Kalchbrenner,
\newblock ``A spectral energy distance for parallel speech synthesis,''
\newblock in {\em Proc. NeurIPS}, 2020.

\bibitem{KHeCVPR2016}
Kaiming He, Xiangyu Zhang, Shaoqing Ren, and Jian Sun,
\newblock ``Deep residual learning for image recognition,''
\newblock in {\em Proc. CVPR}, 2016, pp. 770--778.

\bibitem{ljspeech17}
Keith Ito and Linda Johnson,
\newblock ``The {LJ} speech dataset,''
  \url{https://keithito.com/LJ-Speech-Dataset/}, 2017.

\bibitem{DPKingmaICLR2015}
Diederik~P. Kingma and Jimmy Ba,
\newblock ``Adam: A method for stochastic optimization,''
\newblock in {\em Proc. ICLR}, 2015.

\bibitem{XMaoICCV2017}
Xudong Mao, Qing Li, Haoran Xie, Raymond~Y.K. Lau, Zhen Wang, and Stephen~Paul
  Smolley,
\newblock ``Least squares generative adversarial networks,''
\newblock in {\em Proc. ICCV}, 2017, pp. 2794--2802.

\bibitem{ALarsenICML2016}
Anders Boesen~Lindbo Larsen, S{\o}ren~Kaae S{\o}nderby, Hugo Larochelle, and
  Ole Winther,
\newblock ``Autoencoding beyond pixels using a learned similarity metric,''
\newblock in {\em Proc. ICML}, 2016, pp. 1558--1566.

\bibitem{ABaevskiNeurIPS2020}
Alexei Baevski, Henry Zhou, Abdelrahman Mohamed, and Michael Auli,
\newblock ``wav2vec 2.0: A framework for self-supervised learning of speech
  representations,''
\newblock in {\em Proc. NeurIPS}, 2020.

\bibitem{MHeuselNIPS2017}
Martin Heusel, Hubert Ramsauer, Thomas Unterthiner, Bernhard Nessler, and Sepp
  Hochreiter,
\newblock ``{GANs} trained by a two time-scale update rule converge to a local
  {Nash} equilibrium,''
\newblock in {\em Proc. NIPS}, 2017.

\bibitem{MBinkowskiICLR2020}
Miko{\l}aj Bi{\'n}kowski, Jeff Donahue, Sander Dieleman, Aidan Clark, Erich
  Elsen, Norman Casagrande, Luis~C. Cobo, and Karen Simonyan,
\newblock ``High fidelity speech synthesis with adversarial networks,''
\newblock in {\em Proc. ICLR}, 2020.

\bibitem{CYuIS2020}
Chengzhu Yu, Heng Lu, Na~Hu, Meng Yu, Chao Weng, Kun Xu, Peng Liu, Deyi Tuo,
  Shiyin Kang, Guangzhi Lei, Dan Su, and Dong Yu,
\newblock ``{DurIAN}: Duration informed attention network for multimodal
  synthesis,''
\newblock in {\em Proc. Interspeech}, 2020, pp. 2027--2031.

\bibitem{PGuoICASSP2021}
Pengcheng Guo, Florian Boyer, Xuankai Chang, Tomoki Hayashi, Yosuke Higuchi,
  Hirofumi Inaguma, Naoyuki Kamo, Chenda Li, Daniel Garcia-Romero, Jiatong Shi,
  Jing Shi, Shinji Watanabe, Kun Wei, Wangyou Zhang, and Yuekai Zhang,
\newblock ``Recent developments on {ESPnet} toolkit boosted by conformer,''
\newblock in {\em Proc. ICASSP}, 2021, pp. 5874--5878.

\bibitem{AGulatiIS2020}
Anmol Gulati, James Qin, Chung-Cheng Chiu, Niki Parmar, Yu~Zhang, Jiahui Yu,
  Wei Han, Shibo Wang, Zhengdong Zhang, Yonghui Wu, and Ruoming Pang,
\newblock ``Conformer: Convolution-augmented transformer for speech
  recognition,''
\newblock in {\em Proc. Interspeech}, 2020, pp. 5036--5040.

\bibitem{THayashiICASSP2020}
Tomoki Hayashi, Ryuichi Yamamoto, Katsuki Inoue, Takenori Yoshimura, Shinji
  Watanabe, Tomoki Toda, Kazuya Takeda, Yu~Zhang, and Xu~Tan,
\newblock ``{ESPnet-TTS}: Unified, reproducible, and integratable open source
  end-to-end text-to-speech toolkit,''
\newblock in {\em Proc. ICASSP}, 2020, pp. 7654--7658.

\end{thebibliography}

\clearpage
\appendix

\section{Detailed analysis}
\label{sec:detailed_analysis}

\subsection{Relationship between MOS and cFW2VD/cFDSD}
\label{subsec:mos_cfw2vd_cfdsd}

\begin{figure}[h]
  \centerline{\includegraphics[width=\columnwidth]{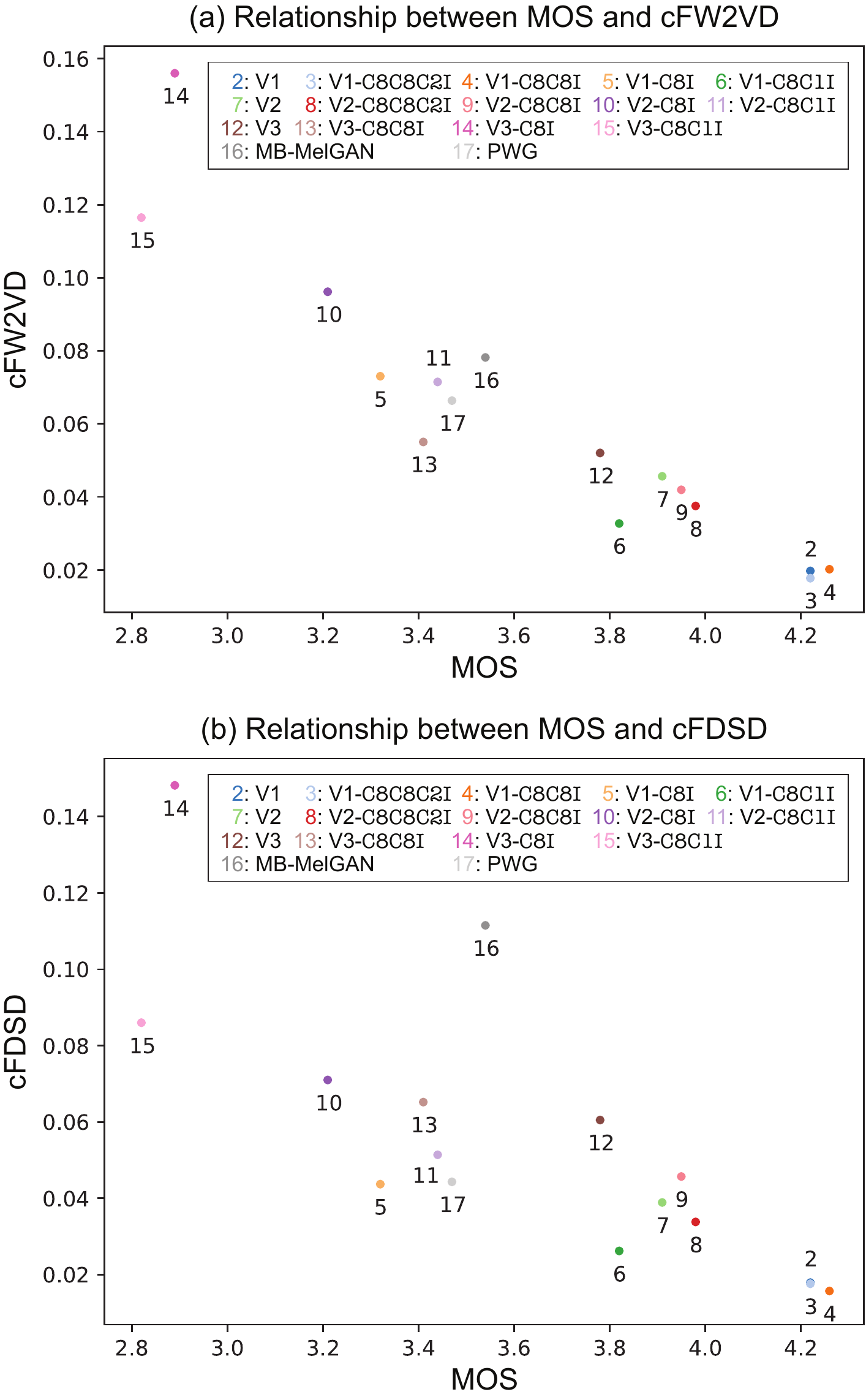}}
  \vspace{-3mm}
  \caption{Relationship between MOS and cFW2VD (a) and that between MOS and cFDSD (b).
    The corresponding MOS and other scores are presented in Table~\ref{tab:result_vocoder}.
    The number under the marker corresponds to the number (No.) in Table~\ref{tab:result_vocoder}.
    The larger the value of the MOS, the better.
    The smaller the value of cFW2VD/cFDSD, the better.
    The MOS has a higher correlation with cFW2VD than with cFDSD (Spearman's rank correlation of $-0.93$ and $-0.83$, respectively).}
  \label{fig:mos_cfw2vd_cfdsd}
\end{figure}

In Fig.~\ref{fig:mos_cfw2vd_cfdsd}, we plot the relationship between the MOS and cFW2VD and that between the MOS and cFDSD.
We used the models listed in Table~\ref{tab:result_vocoder}.
We found that the MOS has a higher correlation with cFW2VD than with cFDSD (Spearman's rank correlation of $-0.93$ and $-0.83$,\footnote{The correlations are negative because the quality improves as the MOS \textit{increases} and cFW2VD/cFDSD \textit{decreases}.} respectively).

\end{document}